\documentclass[prl,aps,amssymb,showpacs,twocolumn]{revtex4}
\usepackage{amsmath}
\usepackage{amssymb}
\usepackage{amsthm}
\usepackage{amsfonts}
\usepackage{enumerate}
\usepackage{latexsym}
\usepackage[dvips]{graphicx}

\newcommand{\beq}{\begin{equation}}
\newcommand{\eneq}{\end{equation}}

\input{epsf}

\begin{document}

\tolerance 10000


\title{Holonomic Quantum Computing Based on the Stark Effect}

\author {B. A. Bernevig and Shou-Cheng Zhang}

\affiliation{  Department of Physics, Stanford University,
Stanford, CA 94305}
\begin{abstract}
\begin{center}

\parbox{14cm}{We propose a spin manipulation technique based entirely
on electric fields applied to acceptor states in $p$-type
semiconductors with spin-orbit coupling. While interesting in its
own right, the technique can also be used to implement
fault-resilient holonomic quantum computing. We explicitly compute
adiabatic transformation matrix (holonomy) of the degenerate
states and comment on the feasibility of the scheme as an
experimental technique.}

\end{center}
\end{abstract}


\maketitle

The physical realization of quantum computing rests on the ability
to reversibly manipulate two level systems called qubits. While
the promise of high computational power is certainly a tantalizing
one, the intrinsic challenges associated with decoherence,
adiabatic evolution, control and noise errors in quantum gate
operations are still to be mastered.

One ingenious way to overcome quantum noise errors is the use of
Non-abelian Holonomic (Geometric) Quantum Computation schemes
\cite{Zanardi}. In these procedures, through the slow tuning of
some external parameters such as applied magnetic or electric
fields, the qubit evolves adiabatically (with constant energy)
around a path that changes its eigenstate from an initial to a
final state. Generically, this quantum evolution is free of
dynamical factors and is geometric in nature, depending only on
the path in parameter space. Geometric holonomy could constitutes
a fault-tolerant way to perform quantum computation \cite{Kitaev},
\cite{Preskill}. Although some experimental systems that would
exhibit such behavior have been proposed \cite{Faoro,Choi},
holonomic quantum computing overall still lacks the variety of
concrete application proposals that conventional quantum
computation enjoys.

In this paper we propose using electric fields to manipulate the
spin of acceptor states in semiconductors with spin-orbit coupling
such as Ge, GaAs and Si. The acceptor impurity ion will bind a
$p$-hole from the spin $3/2$ valence band of the semiconductor
\cite{Pikus1} and the full hamiltonian of the impurity system in
an electric field is given by the linear or quadratic Stark effect
\cite{Pikus2}. Spin-orbit coupling is essential to the existence
of the Stark effect in these semiconductors. There are two doubly
degenerate Kramer states for any value of the electric field and
slowly rotating the electric field induces $SU(2)$ rotations in
the degenerate eigenstates of each energy level. When the electric
field swaps over a cycle and returns to its initial orientation
the holonomy matrix is dependent on the geometry of the swap only.
Consistent with prior theoretical analysis, we take these
holonomies to represent quantum gate transformations
\cite{Pachos1}.

We begin by a short introduction of the main idea of holonomic
quantum computing. We then introduce the hamiltonian of the
acceptor states in an electric field and show how any $SU(2)$
holonomy can be obtained by changing the field's orientation
adiabatically, hence providing the basis for a set of gate
operations. We shortly discuss the viability of such a scheme and
the experimental challenges involved. We close by proposing an
alternative scheme, using external uniaxial strain, which can be
used to achieve spin manipulation in the absence of any external
field. The holonomic qubits discussed here are based on the same
principle as the recently discovered dissipationless spin current
in hole doped semiconductors \cite{murakamiscience}. The
individual qubits can therefore be coupled to each other by a
quantum bus architecture based on the dissipationless spin
current, offering exciting new possibilities towards the
realization of an all solid-state holonomic quantum computer.

In the thesis of holonomic quantum computing, quantum information
is encoded in an $n$-fold degenerate Hilbert space of a
hamiltonian $H_\lambda$ dependent on some external 'control'
parameters (fields) $\lambda$ \cite{Pachos1}. Upon a cyclical
change of these parameters around a loop $C$ during time $T$ such
that $\lambda_{in} = \lambda_{out}$, the system will evolve
between the initial state ($n$ vector) $|\psi \rangle_{in}$ into
$| \psi \rangle_{out} = e^{i\epsilon_0 T} \Gamma(C) |\psi
\rangle_{in}$, where $\epsilon_0$ is the initial eigenvalue
$H_{\lambda_{in}} |\psi \rangle_{in} = \epsilon_0 |\psi
\rangle_{in}$. The first factor is just the dynamical phase, and
will be omitted, while the second factor is the non-abelian
Wilczeck-Zee \cite{WilczekZee} curvature connection (matrix):
\begin{equation}
\Gamma(C) = \textbf{P} e^{\oint_C A^\mu d \lambda_\mu}, \;\;\;
A^\mu_{i j} = \langle \psi^i (\lambda)| \frac{\partial}{\partial
\lambda_{\mu}} | \psi^j(\lambda) \rangle,
\end{equation}
\noindent where $i,j =1,...,n$ and where $\textbf{P}$ represents
the path ordering due to the fact that the gauge connection
$A^\mu$ is now a matrix acting on the degenerate space of
Hamiltonian eigenstates ($\mu$ denotes the different control
parameters). The degenerate Hilbert space of the Hamiltonian
encodes the quantum information where the eigenstates are the
codewords while the non-trivial holonomies associated with it
represent the unitary transformations or 'computations' over the
code. Zanardi and Rasetti \cite{Zanardi} showed that this
prescription is sufficient to implement quantum computation on
single qubit holonomic gates. In subsequent papers \cite{Faoro}
\cite{Choi} \cite{Pachos1} \cite{Pachos2} \cite{Duan} several
schemes for realizing holonomic computation have been proposed.
The schemes involve geometric manipulation of trapped ions
\cite{Duan}, charge pumping within Josephson junction networks
\cite{Faoro}, and Josephson charge qubits \cite{Choi}. Controlled
manipulation of $U(1)$ holonomies (Berry phases) using nuclear
magnetic resonance on a system of weakly coupled $H^1$ and
$C^{13}$ nuclei has been experimentally achieved with great
accuracy by Jones \emph{et al} \cite{Jones}.

It would be of great advantage to have a conventional solid-state
system where holonomic computation can be implemented by using
only electric fields. In this paper we are concerned only with
single qubit holonomy, leaving multi-qubit ones for a later
publication \cite{BernevigChenZhang}. We look at p-type cubic
symmetry semiconductors such as Ge, GaAs and Si. The strong
spin-orbit coupling present in these systems breaks up the valence
bands into two doubly degenerate bands of spin $3/2$ with helicity
$\pm 1/2$ and $\pm 3/2$. The double degeneracy is nothing else
than Kramers degeneracy and is guaranteed by $T$-invariance.
$T$-invariance is maintained even when acceptor impurities (B, Al,
Ga, In) are introduced in the semiconductors. The holes that bind
to these impurities will maintain a certain symmetry subgroup of
the original cubic symmetry of the valence bands they came from.
Let us now consider the effect of an applied external electric
field $E$ on the acceptor-bound hole state. For large electric
fields, the field distortion near the impurity ion can be safely
neglected and the acceptor-hole state has the cubic symmetry of
the crystal $T_d \times I$, giving rise to a quadratic Stark
effect \cite{Pikus2}:
\begin{displaymath}
H_{E^2} = -\frac{  p^2_0}{\varepsilon_i}\{ \alpha E^2 I + \beta
[E_x^2 S_x^2 + E_y^2 S_y^2 + E_z^2 S_z^2  - \frac{5}{4} E^2 I] +
\end{displaymath}
\begin{equation}
 + \frac{2}{\sqrt
3} \delta( E_y E_z\{S_y, S_z\} + E_z E_x \{S_z, S_x\}+  E_x E_y
\{S_x, S_y\}) \},
\end{equation}
\noindent where $\varepsilon_i$ is the ionization energy, $p_0 = e
\overline{r}$ the dipole moment ($\overline{r}$ being the mean
radius of the ground state),and $\overrightarrow{S}$ are the spin-
$3/2$ matrices, describing the valence band states, which
essentially have $P_{3/2}$ character . We have also defined
$\{A,B\} = (AB + BA)/2$. Readers familiar with semiconductor
theory will recognize in the form of $H_{E^2}$ the Luttinger
Hamiltonian structure, with the substitution $\vec{k} \rightarrow
\vec{E}$. This is no coincidence since the symmetry group of both
Hamiltonians is the same. For small applied electric field, we
must take into consideration the local field of the ions, thereby
reducing the symmetry from $T_d \times I$ to $T_d$ and giving rise
to a linear Stark effect \cite{Pikus2}:
\begin{equation}
H_E = \frac{ 2 p \chi}{\sqrt 3}( E_x \{S_y, S_z\} + E_y \{S_z,
S_x\}+ E_z \{S_x, S_y\} ),
\end{equation}
\noindent where $p = e a_B$ with $a_B$ the Bohr radius. The
constants $\alpha, \beta, \delta, \chi, \overline{r}$ are given in
Table 1, although the estimates for  $\chi$ in the literature vary
considerably ($\chi = 0.26$ according to Kopf and Lassman
\cite{Kopf} so the value in Table 1 should be taken as a lower
limit). We want to mention that the donor and acceptor
Hamiltonians and physics are essentially different, with the
donors undergoing only a quadratic Stark shift as opposed to the
acceptor combination of the above linear and quadratic shifts.

\begin{table}
  \centering
  \caption{Values of the coefficients in the linear and quadratic Stark Hamiltonians \cite{Pikus2}}\label{Table 1}
\begin{tabular}{|c|c|c|c|c|c|}
  \hline
   & \;\;\; $\alpha$ \;\;\; & \;\;\;\; $\beta$ \;\;\;\; & \;\;\;\; $ \delta$ \;\;\;\; & \;\;\;\;\;\;\; $\chi$ \;\;\;\;\;\;\; & \;\;\; $\overline{r} (\AA)$ \;\;\;\\
  \hline
  \;\; Ge \;\; & 1 & -0.3 & -0.36 & $0.7 \times 10^{-3}$ & 91 \\
  \hline
   Si & 1 & -0.2 & -0.42 &  1 $\times 10^{-2}$ & 34.4 \\
  \hline
\end{tabular}
\end{table}

\noindent Although for some field $E$ the acceptor Hamiltonian
will be a weighted sum of linear and quadratic Stark effects, we
prefer, without any loss of generality, to work in either of the
two regimes and not in the intermediate one. Each of the
Hamiltonians above has two doubly degenerate Hilbert spaces,
roughly corresponding to values of the $z$-component of the spin
$S_z$ being either $\pm 1/2$ or $\pm 3/2$ (this would be exactly
true if the Hamiltonians were isotropic). The 'control' parameters
are the components of the electric field $\vec{E}$. We must now
show we can achieve 'quantum computations over the code'. These
are represented by $SU(2)$ holonomies over each degenerate Hilbert
space (equivalently, we must show that we can move within an
energy subspace by adiabatically changing $\vec{E}$).

We now prove that such holonomies do indeed exist in our system
and give an explicit generic procedure to calculate them. While we
could just brute-force diagonalize the Hamiltonians above and
treat each of them separately, we prefer to use a more elegant
approach that reveals more of the Hilbert space structure. This
was developed by Demler and Zhang \cite{DemlerZhang} in the
context of the SO(5) theory of high Tc superconductivity, and
extended by Murakami, Nagaosa and Zhang \cite{Murakamiprb} to the
case of hole band in semi-conductors. Readers not interested in
the derivation can jump to the next page where we give the
expression for the $SU(2)$ holonomies. Out of the spin-$3/2$ $J_x,
J_y, J_x$ we can define the new $4 \times 4$ matrices:
\begin{displaymath}
\Gamma^1 = \frac{2}{\sqrt 3} \{S_y, S_z \}, \;\; \Gamma^2 =
\frac{2}{\sqrt 3} \{S_z, S_x \}, \;\; \Gamma^3 = \frac{2}{\sqrt 3}
\{S_y, S_x \}
\end{displaymath}
\begin{equation}
\Gamma^4= \frac{1}{\sqrt 3} (S_x^2 - S_y^2), \;\; \Gamma^5 = S_z^2
- \frac{5}{4} I_{4 \times 4},
\end{equation}
\noindent which satisfy the $SO(5)$ Clifford algebra $\Gamma^a
\Gamma^b + \Gamma^b \Gamma^a = 2 \delta_{ab} I_{4 \times 4}$. In
explicit form, these matrices are:
\begin{equation}
\Gamma^i= \left(%
\begin{array}{cc}
  0 & i \sigma^i \\
 - i \sigma^i  & 0 \\
\end{array}%
\right); \Gamma^4 = \left(%
\begin{array}{cc}
  0 & I \\
  I & 0 \\
\end{array}%
\right); \Gamma^5 = \left(%
\begin{array}{cc}
  I & 0 \\
  0 & -I \\
\end{array}%
\right),
\end{equation}
\noindent where $\sigma^i$, $i=1,2,3$ are the usual Pauli matrices
and $I$ is the identity matrix ( $2 \times 2$ in this case). We
observe that the Hamiltonians $H_E$ and $H_{E^2}$ can now be cast
into a new clean form
\begin{equation}
H_E = d_E^0 I + d_E^a \Gamma^a; \;\; H_{E^2} = d_{E^2}^0 I +
d_{E^2}^a \Gamma^a;  \; a=1,...,5
\end{equation}
\noindent where $I$ is the $4\times 4$ identity matrix and where
the following identities hold:
\begin{displaymath}
d^0_E = d^4_E = d^5_E =0, \; d^1_E = p \chi E_x, \; d^2_E = p \chi
E_y, \; d^3_E = p \chi E_z
\end{displaymath}
\begin{displaymath}
d^0_{E^2} = -\frac{  p^2_0}{\varepsilon_i} \alpha E^2, \;\;\;
d^1_{E^2} = -\frac{  p^2_0}{\varepsilon_i} \delta E_z E_y, \;\;\;
d^2_{E^2} = -\frac{  p^2_0}{\varepsilon_i} \delta E_z E_x
\end{displaymath}
\begin{displaymath}
 d^3_{E^2} = -\frac{  p^2_0}{\varepsilon_i} \delta E_x E_y,
\;\;\; d^4_{E^2} =  -\frac{  p^2_0}{\varepsilon_i} \frac{\sqrt
3}{2} \beta (E_x^2 - E_y^2)
\end{displaymath}
\begin{equation}
 d^5_{E^2} =  -\frac{
p^2_0}{\varepsilon_i} \frac{1}{2} \beta (2 E_z^2- E_x^2 - E_y^2)
\end{equation}
\noindent Since the two Hamiltonians for small and large field
have been brought to the same symbolic form, we can manipulate
them together and only substitute for the values of $d^a$ at the
end of the calculation. The deep physical reason as to why the two
apparently different Hamiltonians are actually very similar is the
unbroken T-invariance of the system that leads to Kramers
degeneracy. The eigenvalues are $\epsilon_\pm  = d^0 \pm d$ where
$d= \sqrt{d^a d^a}$ and they depend on the electric field. For the
linear Stark effect explicit substitution shows us that the split
is independent on the direction of the electric field $\vec{E}$
while for the quadratic case, it strongly depends on its
orientation. The Clifford matrices $\Gamma^a$ have two
eigenvalues, each two-fold degenerate (this is obvious from the
form of $\Gamma^5$). The gauge connection then represents $SU(2)$
adiabatic changes on the two-fold degenerate sub-bands $\pm$ hence
the total gauge group is $SU(2)_- \times SU(2)_+ = SO(4)$.
Identical to the work of Murakami Nagaosa and
Zhang\cite{Murakamiprb}, we can define the projection operators
into the two energy subspaces $H = \epsilon_+ P^+ + \epsilon_-
P^-$:
\begin{equation}
P^+ = \frac{1}{2}(1 + \frac{d^a}{d} \Gamma^a), \;\; P^- =
\frac{1}{2}(1 - \frac{d^a}{d} \Gamma^a).
\end{equation}
\noindent Adiabatic rotation of the field $\vec{E}$ implies moving
within one of the subspaces of energy $\varepsilon$. We can define
a covariant gauge field strength:
\begin{equation}
A_a = i [\frac{\partial P^+}{\partial d^a}, P^+] = i
[\frac{\partial P^-}{\partial d^a}, P^-] = - \frac{1}{2 d^2} d^b
\Gamma^{ab}
\end{equation}
\noindent where $\Gamma^{ab} = \frac{1}{2i} [\Gamma^a, \Gamma^b]$
are the generators of the $SO(5)$ algebra. The field $A_a$ lives
in the space of the $d^a$'s but our control parameters are the
electric field components $E_i$. We hence have to do a 'coordinate
transformation' and obtain:
\begin{equation}
A^i= \frac{\partial d^a}{\partial E_i} A_a = - \frac{1}{2 d^2} d^b
\frac{\partial d^a}{\partial E_i} \Gamma^{ab}
\end{equation}
\noindent which gives us a holonomy computation when the electric
field is varied between $\vec{E}_{initial} = \vec{E}_{final}$ over
an arbitrary (closed) curve $C$
\begin{equation} \Gamma(C) = \textbf{P} \exp(\oint
A^i d E_i) =  \textbf{P} \exp(- \oint  \frac{1}{2 d^2} d^b
\frac{\partial d^a}{\partial E_i} \Gamma^{ab}  d E_i)
\end{equation}
\noindent Let us, without any loss of generality momentarily focus
on the $\epsilon_+$ subspace. By choosing specific rotations
(specific contours $C$) of the field $E_i$ we can change an
initial state $|\psi \rangle_{in} = (1,0,0,0)$ into the degenerate
state within the same energy level, i.e. $|\psi \rangle_{out}
=\Gamma(C) |\psi \rangle_{in} = (0,1,0,0)$. In fact, in the
general case, starting from an arbitrary $|\psi \rangle_{in}$ we
can reach, through carefully choosing the contour $C$, any other
eigenstate within the degenerate subspace by electric field
manipulation. In a physical intuitive picture, the spin within the
$\epsilon_+$ subspace will follow the electric field as it tries
to stay within the energy subspace. We have hence achieved spin
manipulation with electric fields and showed that holonomic
computation is possible in semiconductors with spin-orbit
coupling.

In general, due to the non-abelian nature of $A_i$, the path
ordered integral has to be done numerically, over infinitesimal
segments in parameter space and taking into account that different
components of $A$ do not commute with each other. While this is
more of a nuisance than an intellectual challenge, it is
comforting to know that for certain curves the expression can be
simplified and path ordering can be easily implemented while still
maintaining the full capability to transform the eigenstates into
one another. We give such examples for both the linear and the
quadratic Stark effect below.

For the linear Stark effect, again working in the $\epsilon_+$
energy subspace the expression for the holonomy $\Gamma_E (C)$
becomes particularly simple:
\begin{equation}
\Gamma_E (C) = \textbf{P} \exp(- \frac{1}{2} \oint \frac{1}{E^2}
\epsilon_{ijk} \sigma_k E_j  d E_i)
\end{equation}
\noindent where the $\sigma_k$ are the 3 pauli matrices. In polar
coordinates $\vec{E} = (E \sin \theta \cos \phi, E \sin \theta
\sin \phi, E \cos \theta) $ for contours $C$ which keep constant
the absolute value of the electric field, we find that spherical
triangles between the points $A$ ($\theta =0, \phi=\phi_1$), $B$ (
$\theta = \pi/2, \phi = \phi_1$), and $C$ ($\theta = \pi/2, \phi =
\phi_2$) are particularly easy to path order. Since we are
changing only one angle at a time achieving this technologically
should be easier than trying to implement variations in both
angles (although, as Zee points out, there is a bit of confusion
on how to go 'around the corners' \cite{Zee})

For the case of the quadratic Stark effect things are more
complicated. While finding a nice form for the holonomy factors in
the general case is almost impossible due to the anisotropy in the
Hamiltonian $H_{E^2}$, we can look at the idealized spherical
symmetric situation for which $\beta = \delta / \sqrt{3}$. This
does not introduce large errors, as the anisotropy in these
materials, although significant, is still small enough so that the
spherical approximation works well. In this case we find, in units
of $-\frac{ p^2_0}{\varepsilon_i}$: $d^a_{E^2} \Gamma^a = \beta
({\vec{E}} \cdot {\vec{S}})^2 - 5/4 \beta E^2 I_{4 \times 4}$. The
holonomy structure resides exclusively in the first term. In fact,
with the electric field replaced by a magnetic field, this is
exactly the Hamiltonian studied by Zee \cite{Zee} in explaining a
pioneering experiment by Tycko \cite{Tycko}. The gauge field in
polar coordinates is $A_{\phi} = \cos \theta \sigma_3/2 - \sin
\theta \sigma_1$ and $A_\theta = \sigma_2$. For spherical
triangles starting at $\theta =0$ going to some value $\theta$ on
an arc of fixed $\phi_1$ (which we can choose to be zero for
convenience) then going at fixed $\theta$ on an arc to some
non-zero $\phi$ and then back to the north pole along constant
longitude, the holonomy reads $\Gamma_{E^2} (C) = W_1^{-1} V W$
\cite{Zee} where:
\begin{displaymath}
W_1^{-1} = \exp({-i\theta( \cos \phi \sigma_2 - \sin \phi
\sigma_1)}), \;\;\; W= \exp({i \sigma_2 \theta})
\end{displaymath}
\begin{equation}
V= \exp({-i \frac{\phi }{2} \sigma_3}) \exp({i\frac{\phi}{2}(\cos
\theta \sigma_3 - 2 \sin \theta \sigma_1)})
\end{equation}

We now turn to the problem of the feasibility of the scheme
proposed for spin-manipulation by the Stark effect. We need that
the coherence time of spins of bound holes be larger than the time
in which we can adiabatically rotate the electric field. New
experiments showed that the coherence time is larger than $1 ms$
\cite{Smit}, \cite{Golding}, justifying the use of
acceptor-bound-hole wavefunctions as qubits. It is indeed
difficult to perform experiments which probe non-abelian phase
factors. The original work of Tycko \cite{Tycko} and subsequently
the more complete experiment by Zwanziger, Koenig and Pines
\cite{Zwanziger} on nuclear magnetic quadrupole resonance proved
the existence of the Wilczeck and Zee non-abelian transport of
degenerate states. Instead of rotating the applied fields and
keeping the sample fixed, these experiments kept the applied field
fixed and rotated the sample, which is an equivalent procedure.
The rotation frequencies were of the order of a few kHz (2020 Hz
in Zwanziger \emph{et al}). Imagining an electric-field version of
this experiment, the rotation period of the field is already less
than the bound hole spin coherence time, but further improvement
may be necessary for a realistic measurement. We also need to
guarantee, during the field rotation, that the adiabatic
approximation is accurately maintained and that the acceptor
impurity is not ionized. The ionization energies for acceptor
states are of the order $10 - 60$ meV (see Table 2). The splitting
between the two levels $\epsilon_+$ and $\epsilon_-$ can be
computed from our expressions for their energies (using the
constants in Table 1) and are of the order $10$ meV for Ge in a
field of $10^6$ V/m. Hence the ionization and splitting energies
are roughly the same size and much larger than the applied
electric field frequecy of rotation. The frequencies required for
varying the electric field are hence low enough as to cause
neither ionization of the impurity-hole system nor a breakdown of
adiabaticity. The variation of dynamical phases over the sample
volume, which usually leads to extensive dephasing can be overcome
by an electric field variant of the double-sweep spin-echo
techniques which refocus inhomogeneities in the dynamical phase
but double the effect of the geometric phase \cite{Jones}.

\begin{table}
  \centering
  \caption{Ionization energies in meV for different acceptor impurities (B, Al, Ga) in Si and Ge}\label{Table 2}
\begin{tabular}{|c|c|c|c|}
  \hline
   &\;\; \;\;\; B \;\;\;\;\; & \;\;\;\;\; Al \;\;\;\;\; & \;\;\;\;\; Ga \;\;\;\;\; \\
  \hline
  \;\; Ge \;\; & 10.4 & 10.2 & 10.8 \\
  \hline
   Si & 45.0 & 57.0 & 65.0  \\
  \hline
\end{tabular}
\end{table}

In conclusion, this paper presents a novel way to manipulate the
spins of acceptor impurity-bound hole states in p-type
semiconductors with strong spin-orbit coupling using electric
fields. Depending on its magnitude, the electric field couples
both linearly and quadratically to the spin of the acceptor state
through the Stark effect, but although apparently opposite, the
two effects still maintain the T-invariance of the underlying
semiconductor. The spin manipulation is completely geometric and
realizes, in a practical solid-state system, the theoretical
proposal for holonomic quantum computing. We have obtained an
explicit and general form for the holonomy matrix which transforms
adiabatically transports degenerate eigenstates. While our
analysis is specific to spin $3/2$ it is trivially generalized for
any spin, provided the Stark effect is present. We have also
briefly analyzed the experimental feasibility of the scheme. In a
future work\cite{BernevigChenZhang}, we explore the idea of
different qubits communicating via a bus architecture based on the
dissipationless spin current \cite{murakamiscience}.

In closing, we want to stress that the wide variety of options
available for spin manipulation in semiconductors with spin-orbit
coupling makes them particularly suitable for the realization of
quantum computers. Aside from electric field manipulation, the
spin of the acceptor states (as well as that of the free holes) is
heavily influenced by interior and exterior applied strain on the
material, so that the same analysis presented above can be done
for strain-induced holonomies and spin-manipulation.

In the closing stages of this work, we noticed the independent,
recent work by Yuri Serebrennikov which presents similar ideas to
the ones exposed here \cite{Serebrennikov}.

The authors wish to thank H.D. Chen, R.B. Laughlin and D.I.
Santiago for many stimulating discussions and input. The author
acknowledges support from the Stanford Graduate Fellowship
Program. This work is supported by the NSF under grant numbers
DMR-0342832 and the US Department of Energy, Office of Basic
Energy Sciences under contract DE-AC03-76SF00515.

\end{document}